\documentclass{moriond}

\def\Journal#1#2#3#4{{#1} {\bf #2}, #3 (#4)}

\def\NIMPR{{\em Nucl. Instrum. Methods Phys. Res.} A}

\def\PRL{\em Phys. Rev. Lett.}
\def\PRC{{\em Phys. Rev.} C}
\def\PRD{{\em Phys. Rev.} D}

\def\EPJC{{\em Eur. Phys.} C}
\def\NPA{{\em Nucl. Phys.} A}

\def\be{\begin{equation}}
\def\ee{\end{equation}}
\def\bea{\begin{eqnarray}}
\def\eea{\end{eqnarray}}

\newcommand{\Photo}{}

\begin{document}
\vspace*{4cm}
\title{Recent $J/\psi$ results measured with PHENIX}

\author{ T. Novák (for the PHENIX Collaboration)}

\address{MATE KRC, H-3200 Gyöngyös, Mátrai út 36, Hungary\\
BGE, H-1054 Budapest, Alkotmány út 9-11, Hungary}

\maketitle\abstracts{
Finalized PHENIX results are presented on $J/\psi$
 modification from the RHIC system-size scan, which include $p$+Al, $p$+Au, and 
$^{3}$He+Au (small systems) collisions as well as $AA$ (large systems) collisions at forward and backward rapidity ($1.2<|y|<2.2$).
Comparisons with state-of-the-art theory calculations and LHC data will be discussed. 
The forward rapidity data in small systems are well described by model calculations for all three systems, however, the backward rapidity $p$+Au and 
$^{3}$He+Au data are not reproduced by the shadowing calculation alone. $J/\psi$ $v_2$ meausrements in Au+Au collisions at forward rapidity is consistent with zero.
}

\section{Introduction}

Based on the time scale in which the nuclear collisions occur, the effects that modify the $J/\psi$ production probability can be separated into two categories. Those that occur during the expansion and hadronization of the energy produced in the collision are generally referred to as hot matter effects.
Those that occur on a time scale small than the crossing time of the colliding nuclei are frequently referred to as cold nuclear matter (CNM) effects. 
The latter include modification of the nuclear-parton-distribution functions (nPDFs) in a nucleus \cite{eskola}$^{,}$ \cite{kovarik}, initial state parton energy loss \cite{vitev}, coherent gluon saturation \cite{kharzeev}$^{,}$ \cite{fuji}, breakup of the forming charmonium in collisions with target nucleons \cite{mcglinchey}$^{,}$ \cite{arleo}, and transverse momentum broadening \cite{cronin}.
Here is a short list with definition of CNM effects:
\begin{description}
\item[Gluon Shadowing/Anit-Shadowing:] Modification (suppression/enhancement) of heavy quark crosssection due to modifications of the gluon structure function 
\item[Parton Energy Loss:] The projectile gluon experiences multiple scattering while passing through the target before $J/\psi$ production, reducing the rapidity of the $J/\psi$
\item[Cronin Effect:] Modification of the $J/\psi$ $p_{\mathrm{T}}$ distribution due to multiple elastic scattering of partons
\item[Nuclear Break-Up:] The break up of the bound $J/\psi$ (or precursor state) in collisions with other target nucleons that pass through $J/\psi$ production point
\item[Co-Movers Break-Up:] Final state break up of the $J/\psi$ through interactions with producedpartons 
\end{description}

To explore the effects of energy production on $J/\psi$  yields in light systems, the
PHENIX collaboration has recently measured the modification in three collision systems
involving small projectiles, $p$+Al, $p$+Au and $^{3}$He+Au \cite{acharya}. The $p$+Au and $^{3}$He+Au data,
combined with the earlier PHENIX $d$+Au data \cite{adare}$^{,}$ \cite{adare2}, provide a direct comparison of inclusive
charmonium production for projectiles with one, two and three nucleons on a Au target. 

\section{Experimental Setup}

PHENIX was planned to operate for 10 years, but finally it was running between 2000 and 2016. During the last few years of operation, PHENIX collected data in a wide range of different collision systems. 

The PHENIX Muon Arms are located parallel to the beam pipe, and
measure muons and unidentified charged hadrons at forward ($1.2 < \eta < 2.4$)
and backward ($-2.2 < \eta < -1.2$) rapidity. The Muon Arms compromise
four main components: the FVTX Detector, the Muon Tracker, the Muon
Identifier, and a series of steel hadron absorbers located throughout each
Muon Arm  \cite{akikawa}$^{,}$ \cite{aidala}. The measurements were attainable using the Forward Vertex
Silicon Detector (FVTX), installed in the PHENIX 2012 upgrade, which
covers the pseudorapidity range $1.2 < |\eta| < 2.2$. The FVTX detector, a
precision silicon tracking detector, provides additional space points closest
to the interaction region, supplying the mass resolution necessary to extract
the $\psi$(2S) signal.

\section{Results}
There were significant changes to the PHENIX muon arm configuration and to the PHENIX
simulation framework between the time that the d+Au data set was recorded in 2008 and when
the $p$+Al, $p$+Au, $^{3}$He+Au and their reference $p$+$p$ data were recorded in 2014 and 2015. In
addition, the $d$+Au data set used a different reference $p$+$p$ measurement. Therefore the relative
systematic uncertainties between the modifications measured for $p$+Al, $p$+Au and $^{3}$He+Au are
smaller.

\subsection{Small Systems}

Figure 1 shows the measured rapidity dependence of the modifiation for $p$+Al, $p$+Au and
$^{3}$He+Au, integrated over $p_{\mathrm{T}}$ and for 0-100$\%$ centrality. The data are compared with model
calculations by Shao et al \cite{kusina}$^{,}$ \cite{shao}$^{,}$ \cite{shao2}$^{,}$ \cite{lansberg} that include only CNM effects.

The forward rapidity data are well described for all three systems, as are the backward rapidity
$p$+Al data. However, the backward rapidity $p$+Au and $^{3}$He+Au data are not reproduced by the
shadowing calculation alone. At RHIC energies a significant suppression is expected at backward
rapidity due to breakup of the forming charmonium in collisions with nucleons that have not
yet passed through the production point (commonly referred to as ``nuclear absorption''). An
estimate of this effect has been folded into the shadowing calculation and is included in Figure 1.
Including the absorption estimate brings the backward rapidity calculation into reasonable agreement with the $p$+Au and $^{3}$He+Au data.

\begin{figure}
\centerline{\includegraphics[width=1.0\linewidth]{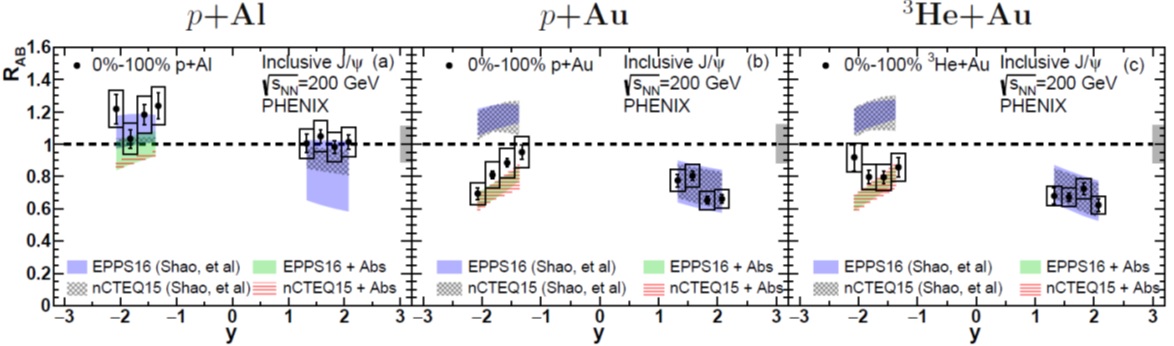}}
\caption[]{Nuclear modification factor of inclusive $J/\psi$ as a function of rapidity for $0-100\%$ $p$+Al (a), $p$+Au (b), and $^{3}$He+Au (c) collisions.
Bars (boxes) around data points represent point-to-point uncorrelated (correlated) uncertainties. The theory bands are discussed in the text.}
\label{fig:fi1}
\end{figure}

Figure 2 shows the $J/\psi$ and $\psi(2S)$ nuclear modification measurements as a function of $\langle N_{coll} \rangle$.
The nuclear modification between the two states follows a similar trend at forward rapidity, with no clear difference
in suppression in the most central collisions. The $\psi(2S)$ shadowing predictions provided by Shao et al. shown
in Figure 2(b) underpredict the suppression at forward rapidity.
Also shown is a comparison to transport-model (TM) predictions for $\psi(2S)$ $J/\psi$ provided by Du and Rapp \cite{du}.

A difference can be seen at backward rapidity between the $J/\psi$ and $\psi(2S)$ nuclear modification, consistent with
a 2.9$\sigma$ effect. The $\psi(2S)$ antishadowing predictions provided by Shao et al. shown in Figure 2(a) do not predict the
suppression. These predictions are purely antishadowing and do not contain any additional CNM effects, such as
nuclear absorption \cite{smith}.

\begin{figure}
\centerline{\includegraphics[width=0.9\linewidth]{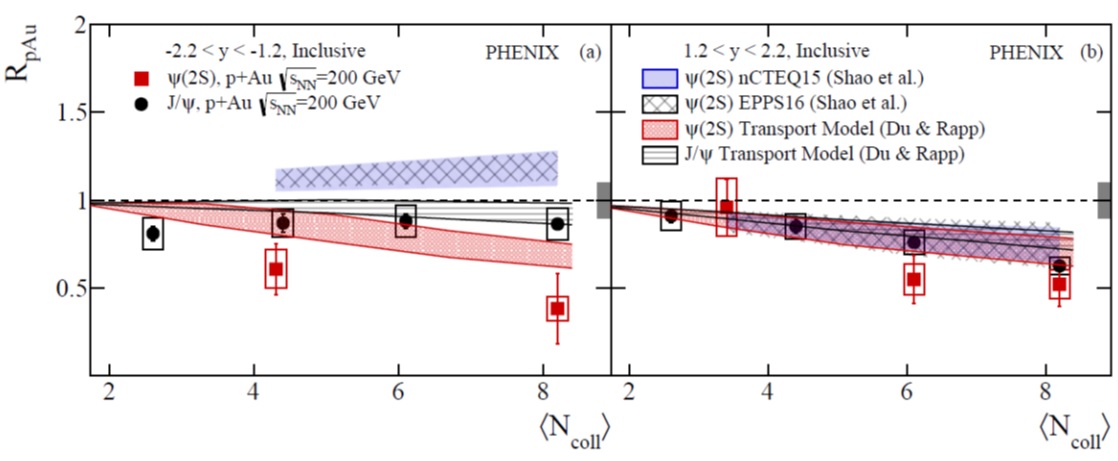}}
\caption[]{Nuclear modification factor of inclusive $J/\psi$ as a function of $\langle N_{coll} \rangle$ for $p$+Au collisions compared with the transport model.
Bars (boxes) around data points represent point-to-point uncorrelated (correlated) uncertainties. }
\label{fig:fig2}
\end{figure}

Figure 2(a) also shows a comparison to TM predictions at backward rapidity, and final-state effects and nuclear
absorption are expected to be important. The TM includes a
nuclear-absorption estimate for both the $\psi(2S)$ and $J/\psi$.
The TMs underpredict the suppression, but describe the
relative modification well and indicate that the $\psi(2S)$
suppression in $p$+Au collisions is consistent with final state effects.

\subsection{Large Systems}
We present a preliminary result for $J/\psi$ $v_2$ using the PHENIX Run14 Au+Au dataset at $\sqrt{s_{NN}}= 200$ GeV \cite{bichon}.
Figure 3 shows the $p_{\mathrm{T}}$-dependent   $J/\psi$ $v_2$. The measurement in this
analysis for PHENIX Run 14 at forward rapidity in a centrality range of 10
- 60$\%$ is shown in red. The measurement made by STAR at mid-rapidity
and in a centrality range of 10-40$\%$ is shown in black. The ALICE result
at forward rapidity in a centrality range of 20-40$\%$ is shown in blue. Boxes
surrounding the data points represent systematic uncertainties.

At LHC energies, a nonzero $v_2$ is observed, which is in line with $J/\psi$
formed by coalescence in the QGP medium, and carrying the azimuthal
anisotropy of the system \cite{acharya2}. At RHIC energies, STAR has measured $v_2$ that
is consistent with zero, but due to limited statistics remains inconclusive \cite{adamczyk}.
With coalescence being the dominant mechanism for nonzero $J/\psi$ $v_2$ it
should follow that systems where fewer $c\bar{c}$ pairs are formed should have a
smaller azimuthal anisotropy.

From the figure we can see the clear nonzero $v_2$ measured by ALICE.
Although the ALICE measurement is at a much higher energy, we know
$v_2$ does not scale with energy for $J/\psi$, so it makes for a good comparison
that the ALICE result which is clearly nonzero is different from our measurement. In our measurement, we see a $v_2$ that is clearly consistent with
zero across all $p_{\mathrm{T}}$ bins. The systematic uncertainties were conservatively
estimated, not taking into account cancellations or correlations of uncertainties from different sources.
Additional data from Run 16 of RHIC will
be included in the final results, and we expect that both statistical and
systematic uncertainties will be significantly reduced.

\begin{figure}
\centerline{\includegraphics[width=0.5\linewidth]{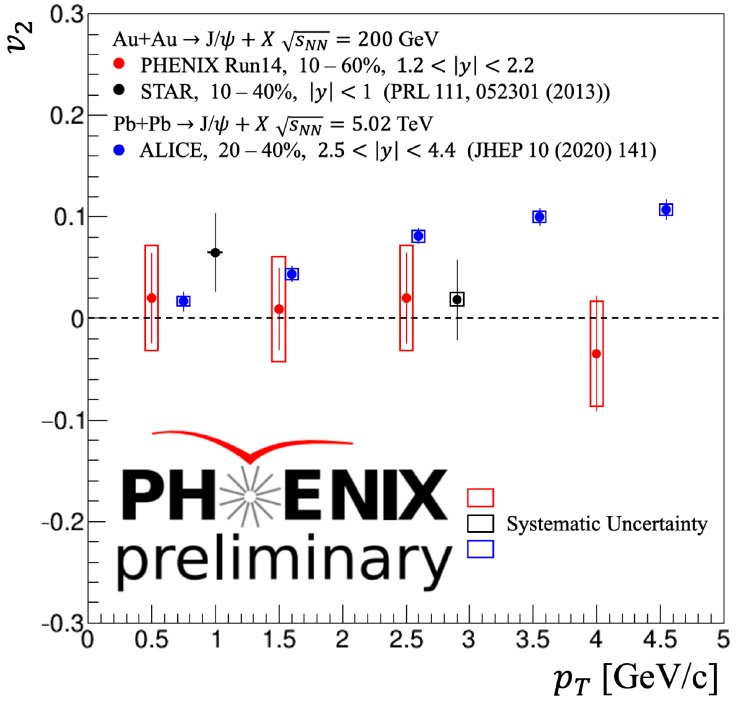}}
\caption[]{Plot of $p_{\mathrm{T}}$ dependent $J/\psi$ $v_2$. The PHENIX result in light gray/red/circle
is compared to STAR in black/star and ALICE gray/blue/square.
 }
\label{fig:fig3}
\end{figure}

\vspace{0.8cm}
 \footnotesize{Partially supported by agencies that support the PHENIX collaboration (https://www.bnl.gov/rhic/funding.php), as well as by the NKFIH grant K133046 (Hungary).}

\end{document}